\documentclass[11pt,twoside]{article}
\usepackage{asp2010}
\usepackage{graphicx}
\resetcounters

\def \hegra {{\em HEGRA}}
\def \hess {{\em HESS}}
\def \magic {{\em MAGIC}}
\def \veritas {{\em VERITAS}}
\def \fermi {{\em Fermi}}
\def \agile {{\em AGILE}}
\def\chandra{{\it Chandra}}
\def \gray {$\gamma$-ray}
\def \grays {$\gamma$-rays}
\aspvolume{in press}
\aspcpryear{2012}
\aspvoltitle{370 Years of Astronomy in Utrecht}
\aspvolauthor{G. Pugliese, A. de Koter, M. Wijburg eds}

\begin{document}

\title{Supernova Remnants as the Sources of Galactic Cosmic Rays}
\author{Jacco Vink,$^1$
\affil{$^1$Anton Pannekoek Institute/GRAPPA, University of Amsterdam,
PO Box 94249, 1090 GE Amsterdam, The Netherlands}
}

\begin{abstract}
           
The origin of cosmic rays holds still many mysteries hundred years after they were first discovered.
Supernova remnants have for long  been the most likely sources of Galactic cosmic rays. I discuss here
some recent evidence that suggests that supernova remnants can indeed efficiently accelerate cosmic rays.
For this conference devoted to the Astronomical Institute Utrecht I put the emphasis on work that was done 
in my group, but placed in a broader context: 
efficient cosmic-ray acceleration and the implications for 
cosmic-ray escape, synchrotron radiation and the evidence for magnetic-field 
amplification, 
potential X-ray synchrotron emission from cosmic-ray precursors,
and I conclude with the implications of cosmic-ray escape for 
a Type Ia remnant like Tycho and a core-collapse remnant like Cas A.
\end{abstract}

%\section{Preamble}
The Astronomical Institute Utrecht had a 370 yr long  history.
But its successes during the more recent past owe much to Marcel Minnaert. 
In this context one  sometimes refers to the ``Minnaert-school'' within
Dutch astronomy.
Its is characterized by its focus on spectroscopy and knowledge of
microphysical processes for understanding astrophysical sources.

Here I discuss cosmic-ray acceleration by supernova remnants (SNRs),
a topic that I started working on as PhD student at SRON, Netherlands Institute for Space Research, 
Utrecht. SRON's history
is connected to Minnaert as well, as SRON was the result of two space science research groups
in Utrecht  and Leiden, both directed by two distinct pupils of Minnaert, respectively Kees de Jager and Henk van de Hulst.

It is also interesting to note that the conference was held not in Utrecht, but at Leeuwenhorst,\footnote{
But there is an Utrecht connection: The name derives from the abbey that was this location until the 16th 
century,
which fell under the jurisdiction of the Bishop of Utrecht.}
Noordwijkerhout.
For me this is also a special place: the high school I attended bears the same name
and is located next to the conference center. 
This high school is
remarkable for having produced two NASA Chandra fellows: Rudy Wijnands and myself.

\section{Introduction}
This year marks the 100th anniversary of detection of cosmic rays by
Victor Hess \citep{hess12}.\footnote{See for \citet{carlson12} for the history behind the discovery of cosmic rays.}
The source(s) of these highly energetic particles, and the way these particles are accelerated, is still
a matter of debate. The energy density of cosmic rays in the Galaxy, about 1~eV\,cm$^{-3}$ has for a long time
been attributed to the power provided by supernovae \citep[e.g.][]{ginzburg64}.
Supernovae are the most energetic events in the Galaxy. Nevertheless, a large fraction, 10-20\%,
of their explosion energy is needed for particle acceleration, in order
 to explain the flux of cosmic rays observed on Earth.
In addition, the lack of spectral features in the cosmic-ray spectrum up to $3\times 10^{15}$~eV suggests
that the sources of cosmic rays should be able to accelerate particles to this energy.\footnote{
Galactic sources should probably even accelerate up  to $3\times 10^{18}$~eV,
as only above this energy the Galaxy becomes transparent for cosmic rays.}
The power provided by supernova explosions may be used to accelerate particles immediately after
the explosion, in the supernova remnant (SNR) phase, or perhaps
in OB associations, due to the combined effects of multiple supernovae and strong
stellar winds \citep{bykov92}.

Most evidence supports now the view that at least part of the cosmic rays originate from the SNR phase.
Since a long time radio observations of SNRs show that relativistic electrons are
present there. But over the last two decades evidence shows that SNRs can accelerate particles
to energies of at least 10~TeV. The first proof was the discovery of X-ray synchrotron radiation from the SN1006 \citep{koyama95}, followed by further evidence for hard X-ray synchrotron emission from Cas A 
\citep{the96,allen97,favata97}. Now for many young SNRs, regions close to the shock front have been identified
whose X-ray emission is dominated by synchrotron radiation. Further proof for acceleration to TeV energies
consists of the detection of TeV gamma-ray emission from many young SNRs 
by Cherenkov telescopes such as \hegra, \hess, \magic\ and \veritas. And since a few years the \fermi\ and \agile\
\gray\ observatories
detected many, both young and mature, SNRs in the GeV \gray\ range.\footnote{ 
See \citet{reynolds08,vink12,helder12} for recent reviews.}
However, there is no conclusive observational evidence yet that SNRs are capable of accelerating particles
up to, or beyond, $3\times 10^{15}$~eV, or that 10\% of the explosion energy is transferred to cosmic rays.
Nevertheless, observations give us several hints that SNRs can indeed accelerate sufficient numbers
of particles to these very high energies.

\section{Efficient cosmic-ray acceleration}\label{sec:acc}

The spectrum and flux of the cosmic rays detected on Earth require that SNRs should be able
to transfer a relatively high fraction of their kinetic energy to cosmic rays, and that they
are capable of accelerating to at least $3\times 10^{15}$~eV.  A few decades this seemed impossible
\citep{lagage83} within the framework of the standard acceleration theory, the so-called diffusive shock
acceleration (DSA) theory.   According to DSA,  particles of sufficient energy can diffusely move with respect
to the overall plasma flow, thereby crossing the shock front. Since there is difference in plasma velocity
between both sides of the shock,\footnote{$v_2\equiv\Delta V=V_1-V_2=(1-1/\chi)V_{\rm s}$, with $\chi$ the shock compression
ratio and $V_{\rm s}$ the shock velocity.}
the particle receives a boost each time it crosses the shock. In fact, it can be shown that $dE/E \approx$~constant. 
Particles advected by the shock-heated plasma that are too far from the shock region 
will be lost from the acceleration process.

The acceleration time scale is given by approximately
%\begin{equation}
$\tau _{\rm acc}= D/V_{\rm s}^2$,
%\end{equation}
with $D$ the diffusion parameter given by $D=\frac{1}{3}\lambda_{\rm mfp} v_{\rm particle}\approx
\frac{1}{3} \eta cE/eB$, in which case the mean free path $\lambda_{\rm map}$
 is assumed to be a factor $\eta$ times the gyro-radius. For a very turbulent magnetic field $\delta B/B\sim 1$,
 $\eta = 1$. The
 length scale over which diffusive transport dominates over advection is given by $l_{\rm diff}=\tau_{\rm acc} V_{\rm s}=D/V_{\rm s}$. For a mean Galactic field of $B\approx 5~\mu$G and $V_{\rm s}=5000$~km\,$^{-1}$, the maximum  energy that can be reached in 500~yr is $E_{\rm max} <\eta^{-1} 6\times 10^{14}$~eV. 
To explain the highest energy Galactic cosmic rays, therefore, 
requires magnetic fields much higher than the mean Galactic field, 
and $\eta$ close to one.
 
 If SNRs are indeed very efficient in converting kinetic energy to cosmic-ray energy, one can no longer
 treat cosmic rays as test particles, but one has to consider the back-reaction of the cosmic rays on the shock
 dynamics itself  \citep{malkov01}.  It turns out
 that in general, cosmic ray acceleration can indeed be very efficient, although critical ingredient into the theory
 that is not well constraint is the injection of particles into the DSA process. Recent results show that easily more
 than 50\% of the pressure can be cosmic ray pressure \citep[e.g.][]{blasi05,vladimirov08}.
The result of this is that cosmic rays streaming ahead of the shock will start compression and heating
the plasma in a so-called cosmic-ray precursor.
For a normal, strong shock one expects a shock compression ratio
of   $\chi=4$ (for $\gamma=5/3$). But the increased compression and change of velocity induced by
the cosmic-ray precursors makes that for an efficiently accelerating shock the gas-shock
compression will be $\chi < 4$. But the combination of shock-compression and cosmic-ray precursor
compression will result in $\chi_{\rm total}>4$. The post-shock plasma temperature will be lower than expected as
the gas is heated by a low Mach number shock.

\begin{figure}
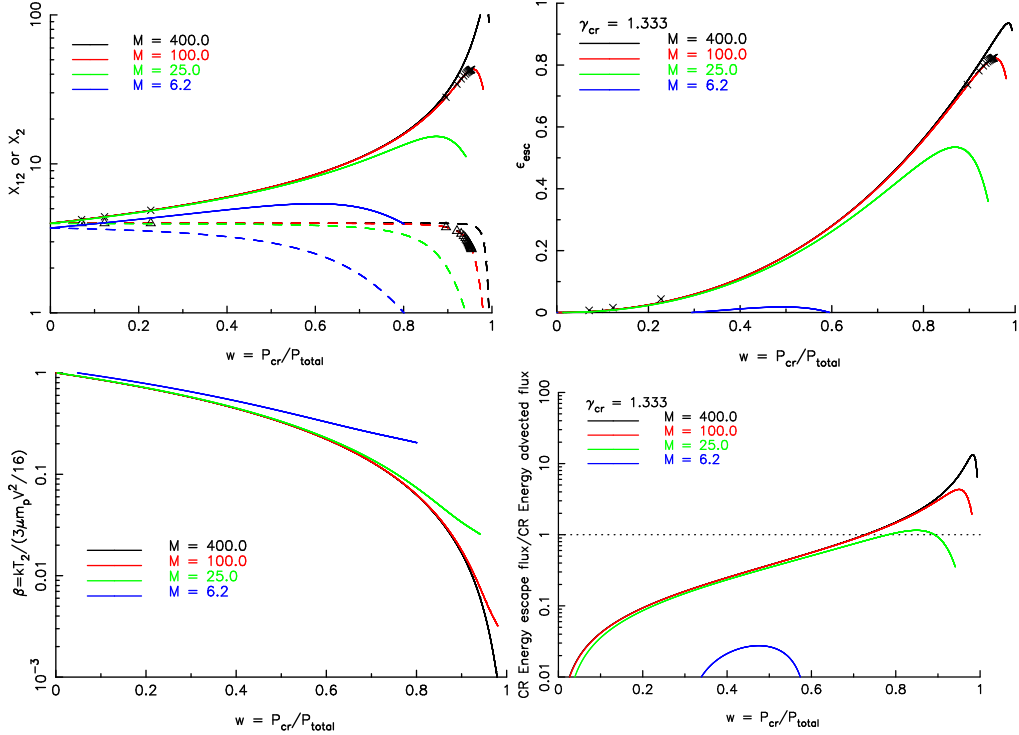

\centerline{
\includegraphics[trim=0 0 0 -20,clip=true,angle=-90,width=0.5\textwidth]{hugoniot_chi12_370yr.ps}
\includegraphics[trim=0 0 0 0,clip=true,angle=-90,width=0.5\textwidth]{hugoniot_eps_370yr.ps}
}
\centerline{
\includegraphics[trim=0 0 0 0,clip=true,angle=-90,width=0.5\textwidth]{hugoniot_beta_370yr.ps}
\includegraphics[trim=0 0 0 -50,clip=true,angle=-90,width=0.5\textwidth]{hugoniot_advected_370yr.ps}
}
\caption{
Shock relations for a 
mixture of hot plasma, $\gamma=5/3$ and cosmic rays, $\gamma=4/3$,
as a function of $w$, i.e. partial cosmic-ray pressure.
Shown are
the total ($\chi_{12}$) and gas-shock ($\chi_2$) compression (top, left);
the energy-escape flux, $\epsilon$ (top right);
the attenuation of the post-shock temperature
(bottom,left); and the ratio of the energy flux escaping 
and the flux advected with the plasma flow (bottom right). 
Mach numbers ($M$) are distinguished by different colors (online edition only).
Crosses indicate calculations based on the model of \citet{blasi05} for
$M=100$. The downsloping parts should be considered unstable solutions.
\label{fig:hugo}
}
\end{figure}

These effects are indicated by numerical acceleration models, but recently \citet{vink10a}
calculated the effects just assuming thermodynamical principles alone, using a two component
fluid: plasma and cosmic rays. This approach can reproduce some of the
results of more complicated calculations. \citet{vink10a} calculated the thermodynamics in three regions: far ahead
of the shock ($0$), in the precursor, just before the plasma enters the shock ($1$) and in the shock-heated region
($2$). The results are parameterized in terms of the particle pressure in cosmic-rays in region 2, $w\equiv P_{\rm cr}/P_{\rm total}$. It is assumed that the cosmic-ray pressure is constant across the shock (from 1 to 2). 
The conservation equations to be solved are just the usual shock equations for mass conservation and pressure equilibrium in each region
$[\rho v], [P+\rho v^2]$.\footnote{The brackets indicate conserved quantities. Velocities are evaluated
in the frame of the shock.} But for the energy flux it assumed that some fraction,
$\epsilon$, of the free energy can be carried away in the form of escaping cosmic rays: 
$(P_2+u_2+\frac{1}{2}\rho_2v_2^2)v_2=(P_0+u_0+(1-\epsilon)\frac{1}{2}\rho v_0^2)v_0$.

The physical reason that energy escape is necessary for efficient cosmic-ray acceleration is that
a shock implies a rapid change in velocity across a region. This can be either accomplished by increasing
the entropy, as is the case in a strong non-cosmic-ray accelerating shock. Or energy has to leak out of the system, 
if cosmic rays are accelerated, because the cosmic-ray acceleration does not result in a big jump in entropy.

The resulting equations for the shock compression ratio, the dependence of cosmic-ray pressure on cosmic-ray escape, and  the post-shock temperature are graphically represented in Fig.~\ref{fig:hugo}. Noteworthy
is that for infinite Mach number the relation between partial cosmic-ray pressure on the one hand, and
the overall compression ratio $\chi_{12}\equiv \chi_{\rm total}$ and main shock compression ratio $\chi_2$, on the other hand,
 is given by $w\approx (\chi_{12}-\chi_2)/(\chi_{12}-1)$. 
 Another interesting result that is usually taken from numerical
simulations, but can be analytical derived, is that the maximum overall compression ratio is reached for
a compression of $\chi_2=\gamma_{\rm gas}/(\gamma_{\rm gas}-1)=2.5$ at the main shock.
Finally,  it turns out that  for Mach numbers $M<6$ efficient cosmic ray acceleration cannot be sustained (blue
lines in Fig.~\ref{fig:hugo}). On face value this means that intra-cluster shocks cannot be efficient particle
accelerators.

The calculated plasma temperatures can be compared to measured plasma temperatures
to infer the cosmic-ray pressure, if the shock velocity is known. This has been investigated now
using optical H$\alpha$ spectroscopy, which
indicates that indeed cosmic-ray pressure is contributing substantially to
the overall pressure in some young SNRs, with $w \gtrsim 20$\% \citep{helder09,helder10}.

\begin{figure}
\centerline{
\includegraphics[trim=30 30 20 20,clip=true,width=0.415\textwidth]{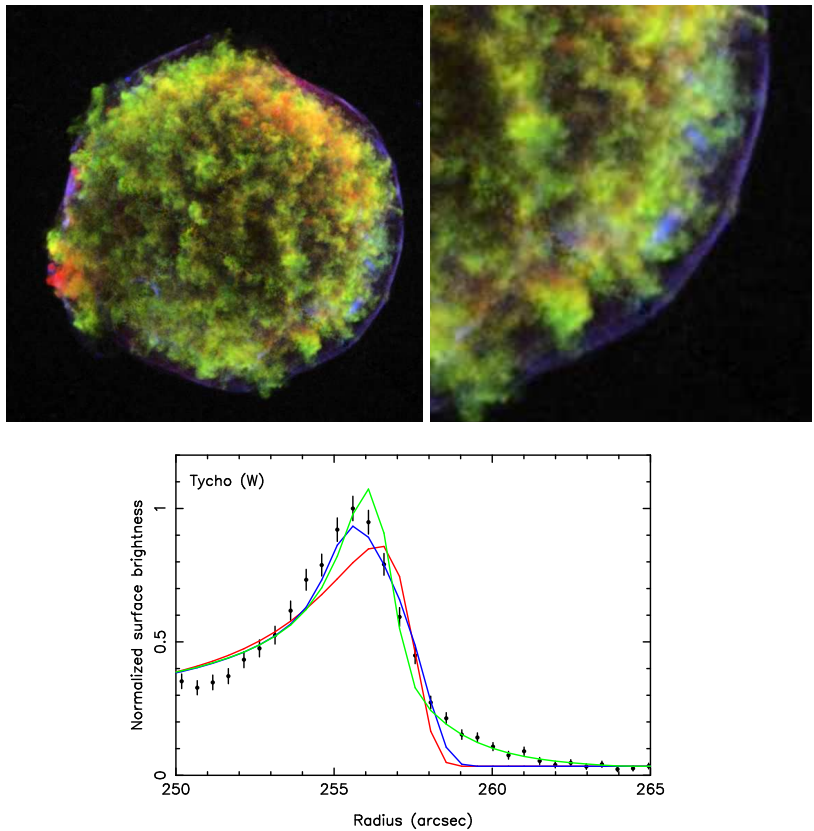}
\includegraphics[trim=320 60 20 260,clip=true,width=0.38\textwidth]{tycho_chandra.ps}}
\centerline{
\includegraphics[trim=0 0 0 -20,clip=true,angle=-90,width=0.55\textwidth]{tycho_rim_west_3models.ps}
}
\caption{
Top:
X-ray image of Tycho's SNR, as observed by \chandra\ in 2009. 
Red indicates Fe-rich plasma,
green Si-rich plasma, and blue strong continuum
emission (X-ray synchrotron radiation).
Right: a detail of the Western shock region.
Bottom panel: Emission profile from a 30\deg\ segment of the Western shock 
region in the 4.3-5.2 keV band. 
The blue line shows a emission from
a uniformly emitting shell with thickness $l$. 
The red and line show  models in which the emissivity falls of behind the shock as $\exp( -(r_0-r)/l)$, but the green line has an additional precursor component. The models take into account spherical projection and instrumental resolution.
(See the online edition for colors.)
\label{fig:tycho}
}
\end{figure}

\section{X-ray synchrotron radiation and evidence for  magnetic field amplification}
\label{sec:synchrotron}

As explained above, the requirement that SNRs should be able
to accelerate above $3\times 10^{15}$~eV in order to explain the cosmic
ray spectrum on Earth, requires both magnetic fields that are
larger than the mean Galactic magnetic field and that the magnetic fields
are highly turbulent (i.e. $\eta \approx 1$).

The detection and properties of the X-ray synchrotron emission from young SNRs
has shown that this may indeed be the case. First of all, the maximum energy
that electrons can be accelerated to is limited by radiative (synchrotron)
losses, which have a typical time scale of $\tau_{\rm loss}=636/B^2E$~s. 
Equating
$\tau_{\rm loss}$ with the typical acceleration time scale $\tau_{\rm acc}$ 
(\S~\ref{sec:acc}) shows that for electrons 
           $E_{\rm max}^2 \propto V_{\rm s}^2/(\eta c eB)$.
The resulting typical maximum photon energy turns out to be
independent of magnetic field \citep[e.g.][]{aharonian99}:
$h\nu=7.4 E^2B~{\rm keV} \approx 
1.4 \eta^{-1}(V_{\rm s}/5000 {\rm km\,s^{-1})^2}$~keV.
Since all young SNRs with  $V_{\rm s}\approx 2000-6000$~km/s
show evidence for X-ray emission ($h\nu \sim 0.1-10$~keV), the magnetic field
must be rather turbulent, i.e. $\eta < 10$. This solves one of the problems for
cosmic-ray acceleration noted by \citet{lagage83}.

The X-ray synchrotron emission from many young SNRs is clearly confined to regions
close to the shock front \citep[e.g.][]{vink03a,bamba05,berezhko03b,helder12}. This is
in particular true for Cas A, Tycho's SNR and Kepler's SNR. Fig.~\ref{fig:tycho} shows
the \chandra\ X-ray image of Tycho's SNR (SN 1572). It beautifully shows the hot thermal emission from
supernova ejecta, shown as a fluffy structured colored red and green in the picture. It
also shows a very narrow, purplish blue, filament
forming the outer boundary of the SNR, which is caused by X-ray synchrotron radiation. 
The thinness of this region can be understood in two ways:
one by noting that as the plasma flow is advected away from the shock with a velocity 
$\Delta v=V_{\rm s}(1-1/\chi)$, carrying away cosmic-ray electrons. Because the electrons suffer radiative
losses, they will after traveling a distance $l=v_2\tau_{\rm loss}$ no longer emit X-rays.

Another way to look at it is, to say that only electrons that remain within a diffusion length $l_{\rm diff}$
of the shock (Sect.~\ref{sec:acc}) will be able to compensate radiative losses with shock acceleration
energy gains. Combining these two ideas in fact shows that the physical widths of the X-ray synchrotron
emitting filaments depend on the average magnetic field strengths behind the shock \citep[e.g][]{berezhko04a,vink04d,parizot06,vink12}:
\begin{equation}
B \approx  %\Bigl(\frac{c}{3e}\Bigr)^{1/3}l_{adv}^{-2/3} =
26 \Bigl(\frac{l_{\rm  synchr}}{1.0\times10^{18} {\rm cm}\sqrt{2}}\Bigr)^{-2/3}\eta^{1/3}\Bigl(\chi_4-\frac{1}{4}\Bigr)^{-1/3}
\ 
{\rm \mu G},\label{eq:bfield}
\end{equation}
with $l_{\rm synchr}$ the width of the synchrotron filaments and $\chi_4$ the overall compression ratio in units of $\chi=4$. Interestingly, this equation is independent of $V_{\rm s}$, but it assumes that
the synchrotron emission is observed with photon energies near the maximum energy.

As an example, Fig.~\ref{fig:tycho} (lower panel) shows
the X-ray synchrotron surface brightness in the Western region. The actual width depends on how
the emissivity falls off with distance to the shock. Acceleration theory predicts an exponential fall-off,
but as the figure indicates this does not give a very good fit. A uniform shell with a sharp edge, provides
a better fit, as it falls-off less sharply from the maximum to larger radii. The exponential model gives a width
of $l=1.2$\arcsec\ corresponding with $l=5.4\times 10^{16}$~cm, and implying a magnetic field
of $B\approx 230~\mu$G. On the other hand, for a uniform emissivity, $l=3.3$\arcsec, which implies $B\approx 100~\mu$G.

During the conference, Colin Norman reminded me of
the effort of Bram Achterberg to use radio-synchrotron radiation
from Tycho's SNR to detect radio synchrotron emission from the shock precursor \citep{achterberg94}. In principle, in X-rays this should be easier to do, as the shock precursor length should
be larger in X-rays. On the other hand, the emissivity scales  as $S\propto B^{\alpha+1}$,
with $\alpha$ the energy flux index. In X-rays $\alpha\approx 2$, whereas in the radio $\alpha\approx 0.6$. 
Since the shock compression makes that in the precursor the magnetic field is lower by a factor $\sim 3$ for
a compression factor of 4, one does not expect to detect the X-ray precursor. But if the shock is efficiently
accelerating,  the main shock may have a compression factor of 2.5 (\S~\ref{sec:acc}), 
which makes that a turbulent upstream
magnetic field is on average only a factor 1.9 weaker, and the synchrotron emissivity only 14\% of the
post-shock emissivity. Indeed, allowing for precursor emission that is 14\% weaker does improve the fit, and results
in an even smaller width in the post-shock region $l=0.4$\arcsec (implying $B \approx 480\, \mu$G), 
and a precursor length of $1.8$\arcsec.
I cannot claim that this is indeed a detection of the precursor, but it is a hint that needs to
be further explored \citep[see also][]{bamba05}.

It turns out that applying Eq.~\ref{eq:bfield} to X-ray synchrotron radiation from young SNRs implies that all
these SNRs have magnetic fields close to the shock front that are substantially larger than the mean Galactic field
\citep{bamba05,voelk05,helder12}. This implies that protons and other atomic nuclei can be accelerated to energies close to $3\times 10^{15}$~eV.  Moreover,
there seems to be a trend that SNRs in lower density environments have lower magnetic fields. The correlation with
shock velocity indicates that $B^2\propto \rho V_s^\beta$ , with $\beta=2-3$ \citep{voelk05,vink08d,helder12}. This is in agreement with recent theories about magnetic field amplification due to cosmic ray induced turbulence \citep{bell04}. 
\citet{helder08} reported evidence for SNR Cas A that also the reverse shock, which heats the ejecta, is 
a dominant source of X-ray synchrotron radiation. Also in the reverse shock region the magnetic field must be
 $100-500\ \mu$G judging from filament widths. What is remarkable about this is that the ejecta are thought to
have a low magnetic field due to the tremendous expansion of the material. That the magnetic field is nevertheless
high suggests that only small seed magnetic fields need to be present to amplify the magnetic fields by large factors.
Or perhaps one should even call it magnetic field creation.

The ideal environment to accelerate to high energies are SNRs that evolve into the winds of their progenitors.
In these winds the density drops of a s $\rho \propto 1/r^2$ \citep[e.g.][]{schure10}.
 And as a result the density is high during the first century of the life of a SNR, when
the shock velocity is also high. The flux of particles entering the shock that can potentially be accelerated is also
large early on, since $F_{\rm cam}\propto \rho 4\pi r^2 V_s$. This is 
$F_{\rm cam}\propto 4\pi (\dot{M}/v_{\rm w})V_s$ for SNRs in a dense wind. 
So more particles are accelerated early on, if SNRs evolve in a dense wind.
The magnetic field amplification ensures a high magnetic field, and particles can be, on average, accelerated for a longer time.

\section{Cosmic-ray escape}
The idea that core collapse SNRs evolving in the dense winds of the progenitor stars are better
at accelerating cosmic rays than Type Ia SNRs is strangely enough not supported by the most recent
\gray\ observations. Cherenkov telescopes show that most young SNRs emit TeV radiation, but for
many it is uncertain whether this is caused by electrons (through inverse Compton scattering) or cosmic-ray
nuclei (pion decay). The GeV telescopes \fermi\ and \agile\ have now greatly increased the number
of SNR \gray\ sources, with many of them being older, core collapse SNRs, often associated
with dense environments, and spectra that cut-off above $\sim 10$~GeV 
\citep[][for a review]{helder12}. Some of these sources have associated
TeV sources that seemed displaced from the SNR itself, suggesting that a nearby cloud is hit by
cosmic rays that have escaped from the SNR \citep[e.g.][]{aharonian08a}. So clearly there is strong connection
between core-collapse supernovae and cosmic ray production. The mature SNRs seem to have lost
the highest energy particles.

When it comes to young SNRs, \gray\ observations do not so clearly indicate that a sizable
fraction of their energy is in the form of cosmic rays. Naively one would expect that a radio-bright and dense
core collapse SNR like Cas A would have a large cosmic ray content, but \fermi\ observations
indicate that less than 4\% of the explosion energy went to cosmic rays \citep{abdo10}.  In contrast, the
Type Ia SNR Tycho seems to have a put a larger fraction of its kinetic energy into accelerating cosmic rays
\citep{acciari11,giordano12}. Given that in \S~\ref{sec:synchrotron} I have argued that core collapse SNRs
evolving in dense winds are probably better accelerators, this contrast between these two young SNRs seems odd.

However, one should also consider that a core collapse SNR like Cas A may have accelerated more
cosmic rays in the first century after the explosion, whereas Tycho may now reach its peak, for the reasons given
in \S~\ref{sec:synchrotron}. The relative lack of \grays 
from Cas A would then imply that 
most cosmic rays have escaped the SNR. As explained in 
\S~\ref{sec:acc} escape is a necessary element for efficiently accelerating shocks.

There can be an additional reason that explains the difference between Tycho and Cas~A. The emission
in pion-decay depends on the local cosmic-ray number density $n_{\rm CR}$, the density of the local
medium, $n_{\rm p}$ and on the emitting volume: $L_{\rm pion} \propto \int n_{\rm CR} n_{\rm p}dV$.
Due to diffusion, cosmic-rays will occupy a region that is larger than the SNR itself, say with radius
$R_{\rm diff}$. The cosmic-ray number
density can be expressed in terms of the ratio of the total number of cosmic rays and the emitting
volume $n_{\rm cr}=N_{\rm cr}/V$. For a Type Ia, like Tycho, which probably evolves in a medium 
with constant density, one sees that the total luminosity of the region containing Tycho only depends
on $N_{\rm cr}$ and $n_{\rm p}$, and is independent of $R_{\rm diff}$:  
$L_{\rm pion} \propto \int n_{\rm CR} n_{\rm p}dV \approx N_{\rm cr} n_{\rm p}$.
This is in contrast to a SNR evolving in a wind:
\begin{equation}
L_{\rm pion} \propto 3N_{\rm cr}/(4\pi R_{\rm diff}^3)\int_0^{R_{\rm diff}}\frac{\dot{M}}{4\pi r^2v_{\rm w}} 4\pi r^2dr \propto \frac{N_{\rm cr}}{R_{\rm diff}^2}\frac{\dot{M}}{v_{\rm w}}.
\end{equation}

The contrast between Tycho and Cas A could, therefore, imply that part of the \gray\ emission
of Tycho may come from a region outside the SNR, whereas for Cas A most \gray\ emission comes from the SNR shell. 
Of course, if in the case of Type Ia SNRs $R_{\rm diff}$ becomes
too large, the emission may fall below the surface brightness limit of the telescope, since the surface brightness
scales with $L_{\rm pion}/R_{\rm diff}^2$. In the future, with a telescope like 
CTA one may want to
search for these cosmic-ray haloes around SNRs.

With this view toward the future I like to conclude my chapter on recent research done by me and my research group at the Astronomical Institute Utrecht. For me this closes off a chapter in my research life of about six years,
which amounts to about 1.6\% of the total history of the astronomical research in Utrecht.  

\acknowledgements 
The closing of the Astronomical Institute Utrecht is a sad loss for the Dutch astronomical landscape. Fortunately many of its former members continue their research elsewhere. 
I would like to
thank all of them for providing a lively atmosphere.  
Special thanks go to all the PhD students and postdocs
who organized many social events, Vanna Pugliese and Marion Wijburg for all the efforts they
made in making this closing conference a success, Frank Verbunt for the many
discussions during lunch and coffee breaks, Bram Achterberg for many discussions
and co-supervising PhD student Klara Schure,
and Christoph Keller for the many hours he put into getting
the best result out of the unfortunate decision
by the Science Faculty.


\begin{thebibliography}{}
\expandafter\ifx\csname natexlab\endcsname\relax\def\natexlab#1{#1}\fi
\expandafter\ifx\csname url\endcsname\relax
  \def\url#1{\texttt{#1}}\fi
\expandafter\ifx\csname urlprefix\endcsname\relax\def\urlprefix{URL }\fi
\providecommand{\eprint}[2][]{\url{#2}}

\bibitem[{{Abdo} et~al.(2010)}]{abdo10}
{Abdo}, A.~A., et~al. 2010, \apjl, 710, L92.

\bibitem[{{Acciari} et~al.(2011)}]{acciari11}
{Acciari}, V.~A., et~al. 2011, \apjl, 730, L20. 

\bibitem[{{Achterberg} et~al.(1994){Achterberg}, {Blandford}, \&
  {Reynolds}}]{achterberg94}
{Achterberg}, A., {Blandford}, R.~D., \& {Reynolds}, S.~P. 1994, \aap, 281, 220

\bibitem[{{Aharonian} et~al.(2008)}]{aharonian08a}
{Aharonian}, F., et~al. 2008, \aap, 481, 401. 

\bibitem[{{Aharonian} \& {Atoyan}(1999)}]{aharonian99}
{Aharonian}, F.~A., \& {Atoyan}, A.~M. 1999, \aap, 351, 330.

\bibitem[{{Allen} et~al.(1997)}]{allen97}
{Allen}, G.~E., et~al. 1997, \apjl, 487, L97.

\bibitem[{{Bamba} et~al.(2005){Bamba}, {Yamazaki}, {Yoshida}, {Terasawa}, \&
  {Koyama}}]{bamba05}
{Bamba}, A., {Yamazaki}, R., {Yoshida}, T., {Terasawa}, T., \& {Koyama}, K.
  2005, \apj, 621, 793.

\bibitem[{{Bell}(2004)}]{bell04}
{Bell}, A.~R. 2004, \mnras, 353, 550

\bibitem[{{Berezhko} et~al.(2003){Berezhko}, {Ksenofontov}, {Ptuskin},
  {Zirakashvili}, \& {V{\" o}lk}}]{berezhko03b}
{Berezhko}, E.~G., et al. 2003, \aap, 410, 189

\bibitem[{{Berezhko} \& {V{\" o}lk}(2004)}]{berezhko04a}
{Berezhko}, E.~G., \& {V{\" o}lk}, H.~J. 2004, \aap, 419, L27

\bibitem[{{Blasi} et~al.(2005){Blasi}, {Gabici}, \& {Vannoni}}]{blasi05}
{Blasi}, P., {Gabici}, S., \& {Vannoni}, G. 2005, \mnras, 361, 907.


\bibitem[{{Bykov} \& {Fleishman}(1992)}]{bykov92}
{Bykov}, A.~M., \& {Fleishman}, G.~D. 1992, \mnras, 255, 269

\bibitem[{{Carlson}(2012)}]{carlson12}
{Carlson}, P. 2012, Physics Today, 65, 020000

\bibitem[{{Favata} et~al.(1997)}]{favata97}
{Favata}, F., et~al. 1997, \aap, 324, L49.

\bibitem[{{Ginzburg} \& {Syrovatskii}(1964)}]{ginzburg64}
{Ginzburg}, V.~L., \& {Syrovatskii}, S.~I. 1964, {The Origin of Cosmic Rays}

\bibitem[{{Giordano} et~al.(2012){Giordano}, {Naumann-Godo}, {Ballet},
  {Bechtol}, {Funk}, {Lande}, {Mazziotta}, {Rain{\`o}}, {Tanaka}, {Tibolla}, \&
  {Uchiyama}}]{giordano12}
{Giordano}, F., et al. 2012, \apjl, 744, L2. 

\bibitem[{{Helder} et~al.(2012){Helder}, {Vink}, {Bykov}, {Ohira}, {Raymond},
  \& {Terrier}}]{helder12}
{Helder}, E.~A., et al., R. 2012, \ssr\ in press, \eprint{arXiv:1206.1593}


\bibitem[{{Helder} et~al.(2010){Helder}, {Kosenko}, \& {Vink}}]{helder10}
{Helder}, E.~A., {Kosenko}, D., \& {Vink}, J. 2010, \apjl, 719, L140.

\bibitem[{{Helder} \& {Vink}(2008)}]{helder08}
{Helder}, E.~A., \& {Vink}, J. 2008, \apj, 686, 1094

\bibitem[{{Helder} et~al.(2009)}]{helder09}
{Helder}, E.~A., et~al. 2009, Science, 325, 719

\bibitem[{{Hess}(1912)}]{hess12}
{Hess}, V.~F. 1912, Physik. Zeitschr., 13, 1084

\bibitem[{{Koyama} et~al.(1995)}]{koyama95}
{Koyama}, K., et~al. 1995, \nat, 378, 255

\bibitem[{{Lagage} \& {Cesarsky}(1983)}]{lagage83}
{Lagage}, P.~O., \& {Cesarsky}, C.~J. 1983, \aap, 125, 249.

\bibitem[{{Malkov} \& {Drury}(2001)}]{malkov01}
{Malkov}, M.~A., \& {Drury}, L. 2001, Reports of Progress in Physics, 64, 429

\bibitem[{{Parizot} et~al.(2006){Parizot}, {Marcowith}, {Ballet}, \&
  {Gallant}}]{parizot06}
{Parizot}, E., {Marcowith}, A., {Ballet}, J., \& {Gallant}, Y.~A. 2006, \aap,
  453, 387. 

\bibitem[{{Reynolds}(2008)}]{reynolds08}
{Reynolds}, S.~P. 2008, \araa, 46, 89

\bibitem[{{Schure} et~al.(2010){Schure}, {Achterberg}, {Keppens}, \&
  {Vink}}]{schure10}
{Schure}, K.~M., {Achterberg}, A., {Keppens}, R., \& {Vink}, J. 2010, \mnras,
  406, 2633. 

\bibitem[{{The} et~al.(1996)}]{the96}
{The}, L.-S., et~al. 1996, \aaps, 120, C357

\bibitem[{{Vink}(2005)}]{vink04d}
{Vink}, J. 2005, in High Energy Gamma-Ray Astronomy,AIP 
Conf. Series, 145, 160. 

\bibitem[{{Vink}(2008)}]{vink08d}
--- 2008, in AIP Conf. Series, 1085, 169. \eprint{arXiv:0810.3680}

\bibitem[{{Vink}(2012)}]{vink12}
--- 2012, \aapr, 20, 49.

\bibitem[{{Vink} \& {Laming}(2003)}]{vink03a}
{Vink}, J., \& {Laming}, J.~M. 2003, \apj, 584, 758.

\bibitem[{{Vink} et~al.(2010){Vink}, {Yamazaki}, {Helder}, \&
  {Schure}}]{vink10a}
{Vink}, J., {Yamazaki}, R., {Helder}, E.~A., \& {Schure}, K.~M. 2010, \apj,
  722, 1727. 

\bibitem[{{Vladimirov} et~al.(2008){Vladimirov}, {Bykov}, \&
  {Ellison}}]{vladimirov08}
{Vladimirov}, A.~E., {Bykov}, A.~M., \& {Ellison}, D.~C. 2008, \apj, 688, 1084.

\bibitem[{{V\"olk} et~al.(2005){V\"olk}, {Berezhko}, \&
  {Ksenofontov}}]{voelk05}
{V\"olk}, H.~J., {Berezhko}, E.~G., \& {Ksenofontov}, L.~T. 2005, \aap, 433,
  229

\end{thebibliography}
\end{document}